\def\mearth{{\rm\,M_E}}
\def\msun{{\rm\,M_S}}

\def\gsim{~\rlap{$>$}{\lower 1.0ex\hbox{$\sim$}}}
\def\lsim{~\rlap{$<$}{\lower 1.0ex\hbox{$\sim$}}}
\def\etal{{\it et al.\thinspace}}
\def\wpm2{W m$^{-2}$}
\def\etal{{\it et al.\thinspace}}
\def\eg{{\it e.g.\ }}

\def\ie{{\it i.e.\ }}

\documentclass[12pt, preprint]{aastex}
\usepackage{graphicx}
\usepackage{natbib,color,lscape}

\title{Tidal Limits to Planetary Habitability}
\author{Rory Barnes\altaffilmark{1,2}, Brian Jackson\altaffilmark{3},
Richard Greenberg\altaffilmark{3}, Sean N. Raymond\altaffilmark{2,4}}

\altaffiltext{1}{Department of Astronomy, University of Washington, Seattle,
WA, 98195-1580}
\altaffiltext{2}{Virtual Planetary Laboratory}
\altaffiltext{3}{Lunar and Planetary Laboratory, University of Arizona, Tucson, AZ 85721}
\altaffiltext{4}{Center for Astrophysics and Space Astronomy, University of
 Colorado, UCB 389, Boulder CO 80309-0389}

\begin{document}
\begin{abstract}
The habitable zones of main sequence stars have traditionally been
defined as the range of orbits that intercept the appropriate amount
of stellar flux to permit surface water on a planet. Terrestrial
exoplanets discovered to orbit M stars in these zones, which are
close-in due to decreased stellar luminosity, may also
undergo significant tidal heating. Tidal heating may span a wide range
for terrestrial exoplanets and may significantly affect conditions
near the surface. For example, if heating rates on an exoplanet are
near or greater than that on Io (where tides drive volcanism that
resurface the planet at least every 1 Myr) and produce similar surface
conditions, then the development of life seems unlikely. On the other
hand, if the tidal heating rate is less than the
minimum to initiate plate tectonics, then CO$_2$ may not be recycled
through subduction, leading to a runaway greenhouse that sterilizes
the planet. These two cases represent potential boundaries to
habitability and are presented along with the range of the traditional
habitable zone for main sequence, low-mass stars. We propose a revised
habitable zone that incorporates both stellar insolation and tidal
heating. We apply these criteria to GJ 581 d and find that it is in
the traditional habitable zone, but its tidal heating alone may be
insufficient for plate tectonics.
\end{abstract}


\section{Introduction}

The discovery of extrasolar planets has made detecting and recognizing
life-bearing planets outside the solar system a real
possibility. Considerations of stellar radiation and climate led to
the definition of a ``habitable zone'' (HZ) as a region around a star in
which a planet with an Earth-like atmosphere could support liquid
water on its surface (Kasting \etal 1993;
Selsis \etal 2007; Spiegel \etal 2009). Close to the star, the
surface temperature is too hot for liquid water; far from the star it
is too cold; but in between is a range of orbits for which liquid
water is stable on the surface. This definition
predicts that habitable orbits lie closer ($\sim 0.1$ AU)
to low-mass stars than to Sun-like stars because low-mass stars are
less luminous. We call this range of orbits the ``insolation habitable
zone'' (IHZ).

Also of critical importance for determining habitability is the
maximum mass that does not accrete a significant hydrogen-rich
envelope. Numerous investigations have modeled this cut-off, and
values range from 0.1 -- 16 Earth masses ($\mearth$) (Pollack \etal
1996; Ikoma \etal 2001; Lissauer \etal
2009). For the expected conditions of most
proto-planetary disks, these results suggest that $10 \mearth$ is a reasonable
approximation for this critical mass, and hence, as is now standard,
we assume that habitable planets must be less massive than this value.

Transit and radial velocity surveys are most sensitive to planets in
small orbits ($\lsim 0.5$ AU), and hence are unlikely to find
terrestrial planets in the IHZs of solar-mass stars. However,
terrestrial-scale planets are readily detectable in the IHZs of
low-mass stars (Mayor \etal 2009). Radial velocity surveys of M stars are complicated by
the presence of stellar activity (West \etal 2004), which can induce
surface fluctuations that erase detectable signals of
terrestrial-scale planets. Nevertheless, the first plausibly
terrestrial-sized exoplanet in the IHZ has just been discovered (GJ
581 d, $\ge 7.1 \mearth$) (Mayor \etal 2009). Analogously, transit
detections are hampered by starspots, which are more frequent and
larger on M stars than on larger stars.  Nonetheless, transit surveys
may be the most effective method to discover terrestrial exoplanets
due to their ability to break the mass-inclination degeneracy of
radial velocity data, and measure the planetary radius directly. With
combined transit and radial velocity data, the unambiguous detection
of a terrestrial planet in the IHZ is feasible in the immediate
future. Not surprisingly, surveys such as ``MEarth'' have begun to
scan M stars for transits (Nutzman \& Charbonneau 2008).

However, planets within a few tenths of an AU from M stars are in a very
different environment than anywhere in the IHZ of our Solar System:
They may be hit with solar flares, bathed in X-rays due to
increased stellar activity, and/or subjected to strong tidal
effects. The first two are certainly concerns for the success of
biology, although they were neglected in the original IHZ
definition. Significant research has since explored these issues (\eg
Lammer \etal 2007; Segura \etal 2009; Zendejas
\etal, in prep.). Here we focus on the third
issue and show that tides may place important constraints on planetary
habitability.

For example, the recently-announced 2 $\mearth$ planet GJ 581 e (Mayor
\etal 2009) may
experience intense tidal heating. Mayor \etal showed that this planet's eccentricity occasionally
reaches values of 0.1. Applying common models of tidal heating (\eg
Peale \etal 1979; Jackson \etal 2008a,b; Barnes \etal 2009), and
assuming the planet to be terrestrial-like, GJ 581 e could have 2
orders of magnitude more tidal heating than Jupiter's volcanic
satellite Io!  Although this
planet is not in the IHZ, similar heating rates on planets in the IHZ
are unlikely to develop life (Jackson \etal 2008a).

However, only in some cases will
the heating be greater than Io's; in others it may lead to
geophysical processes (\eg plate tectonics) that maintain long-term
climate stability, increasing potential habitability (Williams \etal 
1997). Jackson \etal
(2008a) showed that tidal heating in the center of the IHZ could be
substantial for a few illustrative cases. Here we identify the
boundaries of this ``tidal HZ'' (THZ) for a range of stellar and
planetary masses, $M_*$, and $M_p$, respectively, as well as
semi-major axis $a$ and orbital eccentricity $e$. This definition
excludes radiogenic sources of heat, which dominate the heating on the
Earth. Estimating radiogenic heating rates on exoplanets seems a
daunting task as it depends sensitively on the composition of the
planet (itself a result of many different formation pathways
[Raymond \etal 2008]), the age, and the internal structure of the
planet. Therefore, bear in mind that the heating rates we present
here are in addition to any radiogenic heating.

We define a new HZ to take into account tidal heat: Planetary
habitability requires liquid water on the surface {\it and} enough
internal heat to drive plate tectonics, but not so much as to cause
Io-like volcanism. Combined with the other restriction, these
additional requirements significantly reduce the range of detectable habitable
environments around low-mass stars. In $\S$ 2 we describe the IHZ
and THZ. In $\S$ 3 we combine these models to
introduce the ``insolation-tidal HZ'' (ITHZ) and define it for the
cases of 1 and 10 $\mearth$ planets. In $\S$ 4 we discuss the
implications, and suggest directions for future research.

\section{Tidal and Atmospheric Models}

\subsection{The Insolation Habitable Zone}

Kasting \etal (1993) suggested that planetary habitability required
liquid water to be stable on a planetary surface. They made simple
assumptions about the atmosphere, and the planetary orbit, (\eg it was
circular). Selsis \etal (2007) included the effects of cloud cover,
and applied an updated version of the Kasting \etal model to the GJ
581 system (Udry \etal 2007). However, many of the known exoplanets
are on eccentric orbits (Butler \etal 2006), implying rocky exoplanets
will be also. Hence, the effect of eccentricity may be important when
assessing habitability. Williams \& Pollard (2002) pointed out that
planets on eccentric orbits tend to have surface temperatures that
reflect the orbit-averaged stellar flux.

Barnes \etal (2008) modified the Selsis \etal (2007) model to reflect
the Williams \& Pollard (2002) results in order to define an
``eccentric HZ.'' Here we adopt the Barnes \etal definition of the
IHZ, which requires a planet's semi-major axis to lie in the range
$l_{in} < a < l_{out}$, where $l_{in}$ and $l_{out}$ are the inner and
outer edge of the IHZ, respectively. As in Barnes \etal, we assume all
planets have 50\% cloud cover (see Selsis \etal 2007).

\subsection{The Tidal Habitable Zone}

The exact mechanisms of tidal dissipation are poorly understood, but
several quantitative models have been suggested (\eg Goldreich \& Soter
1966; Hut 1981; Levrard \etal 2006; Ferraz-Mello \etal 2008). A conventional model quantifies the tidal heating of a body
as
\begin{equation}
\label{eq:heat}
H = \frac{63}{4}\frac{(GM_*)^{3/2}M_*R_p^5}{Q'_p}a^{-15/2}e^2
\end{equation}
where $G$ is the gravitational constant, $R_p$ is the planetary
radius, and $Q'_p$ is the ``tidal
dissipation function'' which encapsulates the physical response of the
body to tides, including the Love number (Peale \etal 1979; Jackson 2008b).  Note that this
model may break down for large $e$. With the exception
of $Q'_p$, all these quantities can be measured for
exoplanets. However, in lieu of actual mass and radius measurements we
use the scaling relationship $R_p \propto M_p^{0.27}$ (Sotin \etal 2007).

In order to assess the surface effects of tidal heating on a potential 
biosphere, consider the heating flux, $h = H/4\pi R_p^2$, through the 
planetary surface. On Io, $h = 2$ W m$^{-2}$ (from tidal heating) 
(McEwen \etal 2004), resulting in intense global volcanism and a 
lithosphere recycling timescale of 142 -- $3.6 \times 10^5$ years 
(Blaney \etal 1995; McEwen \etal 2004). Such rapid resurfacing on an 
exoplanet probably precludes the development of a biosphere, hence we 
will assume that heating rates larger than $h_{max} \equiv 2$ \wpm2 
result in uninhabitable planets.

Internal heating can also drive plate tectonics. This phenomenon may
enable planetary habitability because it drives the carbonate-silicate
cycle, stabilizing atmospheric temperatures and CO$_2$ levels on
timescales of hundreds of millions of years. Although the processes that drive plate tectonics are not
fully understood on the Earth (Walker \etal 1981; Regenauer-Lieb \etal
2001) an adequate internal heat source is essential. Theoretical
studies of Martian geophysics suggest tectonic activity ceased when
the radiogenic heating flux $h_{rad}$ dropped below 0.04 \wpm2
(Williams \etal 1997). We will therefore assume the
planets have Earth-like weathering and that heating rates
smaller than $h_{min} \equiv 0.04$ \wpm2 result in uninhabitable
planets.

On Earth, the heating comes from the radioactive
decay of U and K and produces a heat flux of $h_{rad\oplus} = 0.08$ \wpm2 (Davies
1999). In the case of exoplanets, the radiogenic heating rates are
unknown, but they might well be
inadequate. Planets significantly smaller than the Earth, which may be
common around M stars (Raymond \etal 2007a), may be
especially susceptible to insufficient heating (the radiogenic heat
flux scales as the ratio of volume to area). In many cases, tidal
heating probably dominates (Jackson \etal 2008a,b), and may be directly estimated from the
mass and orbit of the planet. Such a calculation helps assess a
planet's internal heating and accompanying geophysical processes.

The above examples depend critically on the internal structure and 
composition of the body in question. Without this information, objects 
in the solar system serve as the best guide for hypothesizing the 
internal structure and dynamics of exoplanets. Hence, while explicitly 
acknowledging the wide range of planets likely to be discovered, we 
consider the above definitions of $h_{min}$ and $h_{max}$ to be the 
limits of habitability.

Therefore we define the THZ to be the region around a star for
which $h_{min} < h < h_{max}$ (where $h$ only represents the tidal heating), in this case, enough heat for plate
tectonics, but not so much as to result in extremely rapid
resurfacing. As with the IHZ, this definition makes critical
assumptions about the planetary properties, specifically the planetary
$Q'_p$ value. We assume $Q'_p = 500$, consistent with the values of
most terrestrial-like bodies in the Solar System (Dickey \etal 1994;
Yoder 1995; Mardling \& Lin 2004).

\section{The Insolation-Tidal Habitable Zone}

Habitable conditions may be present on a planet if it lies in {\it
both} the IHZ and the THZ. In other words, $l_{in} < a < l_{out}$ and $h_{min} < h < h_{max}$. When both these requirements are met, a
planet is in the ``insolation-tidal HZ'' (ITHZ). In this section we
present examples of how this refinement modifies our view of
habitability of planets orbiting main sequence stars.

\begin{figure}
\plotone{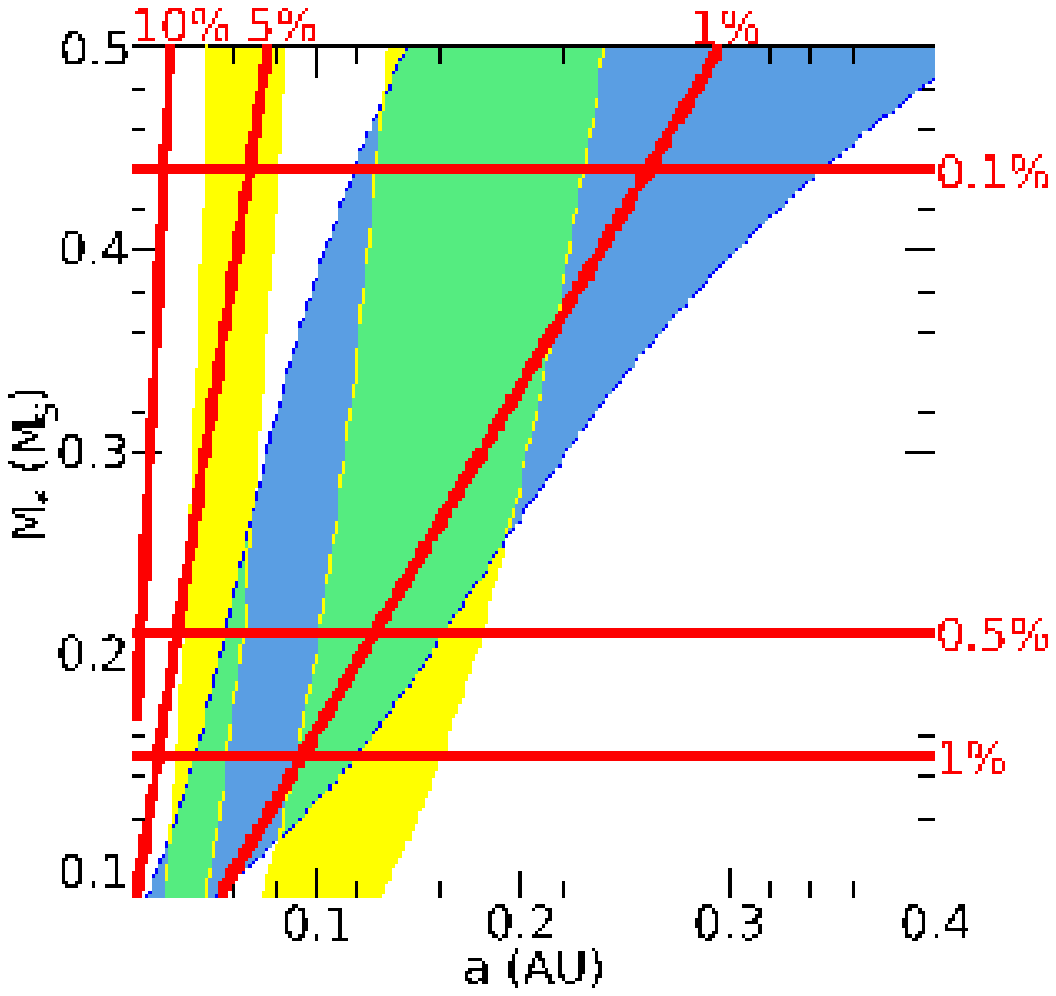}
\caption{Boundaries of habitability for a 10 $\mearth$ terrestrial planet, and
associated transit detectability levels. The blue region represents the IHZ
boundaries for a planet with 50\% cloud cover (the boundaries change
negligibly for $e < 0.5$). The two yellow strips correspond to planets with
favorable amounts of tidal heating if $e = 0.01$ (left strip), or $e = 0.5$ (right strip). Horizontal red lines correspond to different transit depths (with the value of each line to the right). Diagonal red lines represent geometric transit probability (with the value of each line at the top).}
\label{fig:am}
\end{figure}

Figure \ref{fig:am} shows the ITHZ as a function of $M_*$ and $a$ for
a 10 $\mearth$ planet, the most detectable terrestrial-like
exoplanet. The blue shading represents the IHZ (see $\S$ 2.1); the
yellow shading represents the THZ (see $\S$ 2.2). For the latter, two
cases are shown: e = 0.01 (left strip), and e = 0.5 (right strip). The
intersections of the IHZ and the THZs are shaded green and are two
ITHZs. Apparently by chance, the two habitability requirements
significantly overlap in these and many other plausible cases, but the
range of habitable orbits is smaller when considering both tidal and
insolation effects.

In order to assess the likelihood of finding bodies in the ITHZ, we show transit detection probabilities
as the red lines in Fig.\ \ref{fig:am}. The horizontal ones are
geometric transit probabilities (calculated from Barnes [2007]), and the
diagonal ones represent the percentage of starlight blocked during
transit (the ``transit depth'', the ratio of the cross-sectional areas
of the planet and star). Geometric probabilities increase with
decreasing $a$, but transit depths are independent of $a$. Transit
depths of $\sim$ 0.5\% have been detected from the ground (\eg Winn \etal
2008). Transiting terrestrial-scale planets are most likely to be
observed in the lower left of this figure, and hence may experience
significant tidal heating. The recently-announced $\sim 10 \mearth$
planet CoRoT-7 b orbits a $\sim 0.75 \msun$ star and hence has a large
geometric transit probability, but the transit depth was very low,
requiring space-based observations to detect it.

Lower mass planets undergo less tidal heating and block less light,
\ie the yellow strips in Fig.\ \ref{fig:am} move to the left, and the
horizontal red lines move down. The geometric transit
probability scales as the sum of the stellar and planetary radii
(which scales with mass). But since terrestrial planet radii are much
smaller than stellar radii, the diagonal red lines change negligibly
with planetary mass. The IHZ is independent of the planetary
mass and radius (as long as the planet is massive enough to retain an
atmosphere). 

Next we consider the ITHZ for a range of $M_*$, $M_p$, $a$ and
$e$. Figure \ref{fig:ae} shows the two HZ boundaries for 1 and $10
\mearth$ planets (columns) orbiting $0.15$ and $0.25 \msun$ stars
(rows) as a function of $a$ and $e$. The color scheme is the same as
in Fig. \ref{fig:am}.

As stellar mass increases, both the IHZ and THZ move to larger $a$,
and the two HZs move more or less together. However as planetary mass
increases, the THZ may sweep through the IHZ, everything else held
constant. The IHZ moves to larger $a$ at larger $e$ because the
orbit-averaged flux increases with $e$ (Williams \& Pollard 2002). The
tidal heating limits move to larger $a$ with larger $e$ because larger
eccentricity leads to more heating.

We can apply these concepts to GJ 581 d (Mayor \etal 2009), whose mass 
and orbit were recently revised to $M_p = 7.09 \mearth$, $a = 0.2184$ 
AU, and $e = 0.38$. These parameters predict the planet is in the Selsis 
\etal (2007) IHZ. However, using a stellar mass of 0.31 $\msun$ (Udry 
\etal 2007), the tidal heat flux is 0.01 \wpm2 = $0.25h_{min}$. 
Therefore our analysis excludes this planet from the ITHZ. However, if 
the planet has at least 0.03 \wpm2 of radiogenic heating, it could still 
be habitable.

\section{Discussion and Conclusions}

We have shown that consideration of tidal heating and stellar flux are
crucial when assessing planetary habitability. Tidal heating and
detectability scale with stellar and planetary masses, but decrease
with increasing $a$. Hence many planets found in the IHZ may
in fact be uninhabitable due to intense global volcanism and rapid
resurfacing.  Consideration of tidal heating provides an important
refinement in the definition of an HZ.

We have applied our model to the planet GJ 581 d (Mayor \etal
2009). This planet, although in the IHZ, does not reside in the
THZ. However, this planet may still possess plate tectonics through
radiogenic heating. At this point we cannot conclude that GJ 581 d is
habitable. We encourage future work that examines the radiogenic
heating of this planet.

We have assumed that tidal heating can drive plate tectonics,
including subduction. For this possibility to be true, heating has to
occur deep enough inside the planet to drive mantle
convection. No models or observations of tidally-driven subduction
have been made. Such geophysical modeling would be illuminating and
could necessitate a revision of the ideas presented here.

Our choice of $h_{min}$, though rooted in observations in the Solar
System, could be inappropriate in some cases. The process of plate
tectonics is studied by a large and active group of geophysicists, and
as new results come to light, we may need to change our working
definition of $h_{min}$. Similarly, our choice for $h_{max}$ may
require revision. Planets do not recycle their
lithosphere uniformly; ``cool spots'' could persist for long times and
be oases for biogenesis. Furthermore, planets
with unusual compositions or surface features may be able to maintain a biosphere
even with such large heating. Future geophysical and biological modeling could
elucidate the true value of $h_{max}$.

Although we have generally focused on tidal heating, any heating with
the potential to drive plate tectonics may stabilize climates. Therefore the ITHZ is really a subset of a broader
``insolation-geophysical HZ''. It may be that very few planets have $h
< h_{min}$ because radiogenic heating is generally larger than $h_{min}$ (but only for a finite amount of time). However, the distribution of
radiogenic isotopes in the galaxy remains unknown. Although the origin of short-lived radionuclides such as
$^{26}$Al in the Solar System is not fully understood, models suggest
that there should exist a dispersion of several orders of magnitude
in their in protoplanetary disks, and most
planetary systems should actually contain less $^{26}$Al than the Solar
System (Gaidos \etal 2009; see also Gounelle \etal 2009).  If abundances of K and U (the main sources of
internal heat on the Earth) also have a large dispersion, then many
exoplanets may not support radiogenically-driven plate tectonics. 
For such planets, tidal heating may be the only available energy
source (Williams \etal 1997).

\begin{figure}
\plotone{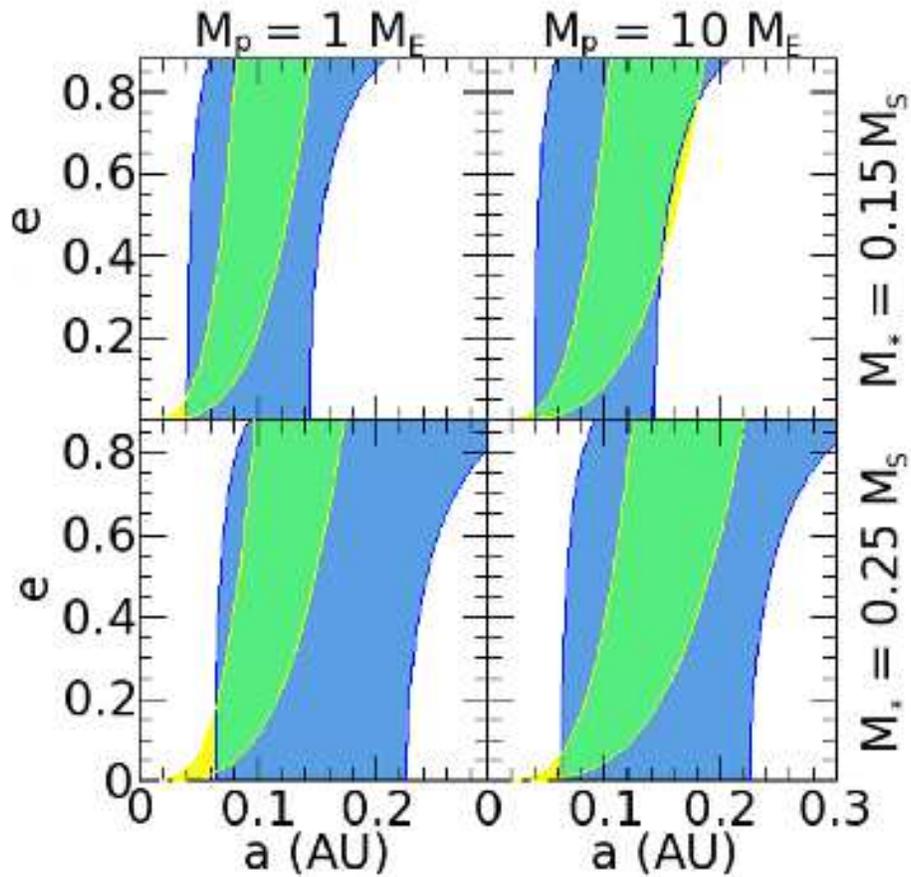}
\caption{Habitability boundaries for a 1 $\mearth$ planet (left
column) and 10 $\mearth$ planet (right column) orbiting a 0.15 $\msun$
star (top row) and 0.25 $\msun$ star (bottom row). This figure uses 
the same color scheme as Fig.\ \ref{fig:am}.}
\label{fig:ae}
\end{figure}

The orbital history of a planet, as well as where it is now, may be of
critical importance when assessing habitability. Most planets inside
0.2 AU have orbits nearly circularized by tides from earlier eccentric
orbits (Jackson \etal 2008c). As $e$ dropped (over billions of years),
so did $a$. In general, $e$ decays more quickly than $a$ (planets tend
to move vertically downward in Fig. \ref{fig:ae}) so a planet can form
interior to the ITHZ on a high-$e$ orbit which
is then circularized by tides into the ITHZ. If we detect a
planet with such a history, it is probably not habitable. The reason
is that large terrestrial planets, \eg $\sim 10 \mearth$, will tend to
hold onto their atmospheres, and hence may be dominated by outgassed
species. Thus, a planet in the ITHZ today could be uninhabitable due
to a volatile past. Spectral analyses of transiting planets could
reveal evidence of extreme volcanism, even though it is not active
today. We encourage atmospheric modelers to consider the evolution of
such a planet.

As tides modify the orbit, the eccentricity may eventually
reach zero, and Eq.\ (\ref{eq:heat}) predicts no heat flux. Without
radiogenic heating, the planet would become
uninhabitable. However, other planets in
the system may perturb the orbit and maintain a nonzero eccentricity
(Mardling 2007), such as in the Galilean satellite system. Models of
the formation of large terrestrial planets in small orbits suggest
such planets may be likely (Raymond \etal 2008). Therefore, despite the persistent damping of $e$ by tides, some
planets may reside in the ITHZ for a very long time.

As discussed in $\S$ 1, tidal effects are but one of several exotic
issues that could affect the habitability of planets orbiting M
stars. For example, the stripping of atmospheres due to massive
stellar winds (ablation) is
also likely a function of $M_*$, $M_p$, and $a$ (Zendejas \etal, in prep.). In principle,
detailed numerical modeling of these phenomena could identify an
``ablation HZ'', which could be compared to the ITHZ presented
here. The identification of an ``Insolation-Tidal-Ablation HZ'' may
guide the wide array of resources that are coming on line to detect
terrestrial mass planets. Other refinements to the limits of
habitability incorporating other stellar and planetary phenomena, such
as obliquity (Williams \& Pollard 2003; Spiegel \etal 2009), the
presence of a large Moon (Laskar et al 1993; 
but see also Barnes \& O'Brien 2002),
or the presence of water on the planet (Kuchner 2004; Raymond \etal
2007b), could be just as useful as the consideration of ablation.

\vspace{0.3cm}
RB and SNR acknowledge funding from NASA Astrobiology Institute's
Virtual Planetary Laboratory lead team, supported by NASA under
Cooperative Agreement No. NNH05ZDA001C. BJ acknowledges support from a
NESSF. RG acknowledges support from NASA's Planetary Geology and
Geophysics program, grant No. NNG05GH65G. We thank Victoria Meadows
and Shawn Domagal-Goldman for enlightening discussions.

\clearpage
\clearpage

\clearpage
\clearpage

\end{document}